\documentclass[sigconf]{acmart}  % 
\usepackage[utf8]{inputenc}
\usepackage{hyperref}  % hyperref should be loaded late
\usepackage{pdflscape}
\usepackage{booktabs}  
\usepackage{array}     % For column formatting
\usepackage{xcolor}    % For cell color
\usepackage{rotating}  % For rotating text
\usepackage{fontawesome5}
\usepackage{threeparttable}
\usepackage{tablefootnote}
\usepackage[most]{tcolorbox}  % Loaded once with the [most] option
\usepackage{balance}
\usepackage{xcolor}
\newtcolorbox{takeawaybox}[1]{
    enhanced,
    colback=white,
    colframe=gray!30!white,
    coltitle=black,
    fonttitle=\bfseries,
    fontupper=\small,
    title={\large\faBook~#1},
    boxrule=1pt,
    sharp corners,
    boxed title style={
        colframe=white, 
        colback=gray!30!white, 
        sharp corners,
        left=1mm,
        right=1mm,
    },
    attach boxed title to top left={
        yshift=-\tcboxedtitleheight/2,
        xshift=1mm,
        yshifttext=-\tcboxedtitleheight/2,
    },
    top=0mm,
    bottom=0mm,
    left=3mm,
}

\usepackage{tikz}
\newcommand{\blackcircle}[1]{%
    \begin{tikzpicture}[baseline=-0.75ex]
        \node[circle, fill=black, inner sep=1.5pt, text=white, text width=0.5em, align=center] {#1};
    \end{tikzpicture}%
}
\usepackage{comment}

\copyrightyear{2024}
\acmYear{2024}
\acmConference[EASE 2024]{The 28th International Conference on Evaluation and Assessment in Software Engineering}{18–21 June, 2024}{Salerno, Italy}
\acmPrice{15.00}
\acmDOI{10.1145/xxxxxxx.xxxxxxx}
\acmISBN{978-1-4503-9053-8/24/06}

\title{Issues and Their Causes in WebAssembly Applications: An Empirical Study}
\author{Muhammad Waseem}
\affiliation{%
  \institution{Faculty of Information Technology, University of Jyväskylä}
  \city{Jyväskylä}
  \country{Finland}
}
\email{muhammad.m.waseem@jyu.fi}

\author{Teerath Das}
\affiliation{%
  \institution{Faculty of Information Technology, University of Jyväskylä}
  \city{Jyväskylä}
  \country{Finland}
}
\email{teerath.t.das@jyu.fi}

\author{Aakash Ahmad}
\affiliation{%
  \institution{School of Computing and Communications, Lancaster University Leipzig}
  \city{Leipzig}
  \country{Germany}
}
\email{a.ahmad13@lancaster.ac.uk}

\author{Peng Liang}
\affiliation{%
  \institution{School of Computer Science, Wuhan University}
  \city{Wuhan}
  \country{China}
}
\email{liangp@whu.edu.cn}

\author{Tommi Mikkonen}
\affiliation{%
  \institution{Faculty of Information Technology, University of Jyväskylä}
  \city{Jyväskylä}
  \country{Finland}
}
\email{tommi.j.mikkonen@jyu.fi}

\renewcommand{\shortauthors}{Waseem et al.}

\begin{document}

\begin{abstract}
WebAssembly (Wasm) is a binary instruction format designed for secure and efficient execution within sandboxed environments - predominantly web apps and browsers - to facilitate performance, security, and flexibility of web programming languages. In recent years, Wasm has gained significant attention from the academic research community and industrial development projects to engineer high-performance web applications. Despite the offered benefits, developers encounter a multitude of issues rooted in Wasm (e.g., faults, errors, failures) and are often unaware of their root causes that impact the development of web applications. To this end, we conducted an empirical study that mines and documents practitioners’ knowledge expressed as 385 issues from 12 open-source Wasm projects deployed on GitHub and 354 question-answer posts via Stack Overflow. Overall, we identified 120 types of issues, which were categorized into 19 subcategories and 9 categories to create a taxonomical classification of issues encountered in Wasm-based applications. Furthermore, root cause analysis of the issues helped us identify 278 types of causes, which have been categorized into 29 subcategories and 10 categories as a taxonomy of causes. Our study led to first-of-its-kind taxonomies of the issues faced by developers and their underlying causes in Wasm-based applications. The issue-cause taxonomies - identified from GitHub and SO, offering empirically derived guidelines - can guide researchers and practitioners to design, develop, and refactor Wasm-based applications.

%Overall, we identified 120 types of issues, which have been categorized into 19 subcategories and 9 categories as a taxonomy of issues. Similarly, we identified 278 types of causes, which have been categorized into 29 subcategories and 10 categories as a taxonomy of causes. Our study led to the first-of-its-kind taxonomies of issues faced by developers and their underlying causes in Wasm-based applications

%esults indicate that predominant issues faced by developers arise from `Infrastructure and Compatibility Aspects' (28.16\%), `Language Features and Documentation Errors' (18.00\%), along with `Code Implementation and Build failures' (13.83\%). Analysing the root-causes indicate that `Syntactic and Semantic Errors' (25.77\%), `Configuration and Compatibility Constraints' (20.1\%), and `Operational Limitations' (12.98\%) are the principal causes of these issues. The study provides a taxonomical classification of issues and their causes, offering empirically derived guidelines, that can guide researchers and developers to systematically design, develop, and refactor Wasm-based applications.
\end{abstract}


\maketitle

\section{Introduction}
\label{Introduction}
WebAssembly (Wasm) as a binary instruction format enhances the performance and security of applications in web-based execution environments \cite{lehmann2020everything}. It serves as a potential compilation target for a variety of programming languages including C, C++, and Rust, marking a significant milestone in web development \cite{haas2017bringing}. Wasm enables developers to employ their chosen programming languages and execute them swiftly in browsers, elevating the overall web experience that can range from gaming to multimedia and scientific simulations \cite{ketonen2022examining}. One of the key features of Wasm is its sandboxed execution environment - a compelling alternative to JavaScript, generally regarded as default language for web applications - offering an efficient and secure web interactions \cite{goltzsche2019acctee}.

Recent research (e.g., \cite{kotilainen2023webassembly}\cite{kotilainen2022proposing}) has shown a significant rise in the utilization of Wasm beyond web browsers. This entails adapting code from various programming languages to operate on a range of devices via the Wasm Interpreter, aiming to establish a unified software architecture for web systems, softwares, and services. 
%This approach can potentially simplify web or networked systems, enhance performance, and lessen the complexity for developers in building sophisticated web or networked systems. 
Despite these advantages, a thorough understanding of the challenges encountered by developers working with Wasm applications is yet to be fully explored. Wasm is a promising technology, however; its ecosystem and associated tools are in a phase of continuous evolution and often regarded as unstable that can impede the development practices for web applications \cite{herrera2018webassembly}. Wasm applications may have an \textit{additional} or \textit{specific} set of \textbf{issues}. Borrowing the idea from \cite{waseem2021nature} \cite{waseem2023understanding}, we define \textbf{issues} in this study as errors, faults, failures, and bugs that occur in Wasm applications and consequently impact their quality and functionality. %Research and development focused on systematically identifying the issues and their root-causes can help researchers and practitioners to systematically tackle the design, development, and refactoring of Wasm applications.
To address this \textit{knowledge gap}, we conducted an empirical study of developer interactions on GitHub and Stack Overflow (SO) concerning Wasm applications. By scrutinizing the information within these exchanges, we aim to identify the common problems faced by developers and the underlying causes of these issues. This initiative will help identify the common issues and their causes, which can also highlight areas requiring further research.

\begin{figure}[!htbp]
\scriptsize
 \centering
 \includegraphics[scale=0.62]{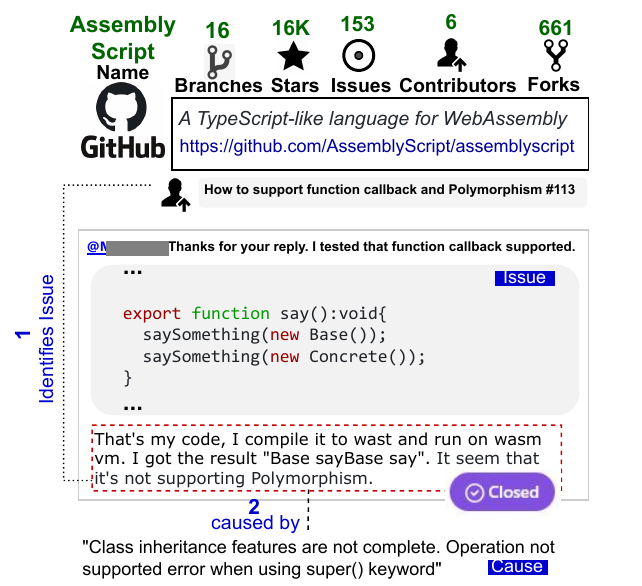}
 \caption{Example: Github based Issue and its Cause in Wasm}
 \label{fig:MotivationExample}
\end{figure}

\textbf{\textit{Motivating Scenario}}: To contextualise the issues and causes in Wasm, we have provided a representative example in Figure \ref{fig:MotivationExample}. The particular example is taken from the Assemblyscript project, an open-source Wasm based project hosted on GitHub (see Table \ref{tab:webassembly-projects}). Figure \ref{fig:MotivationExample} provides essential information about Assemblyscript, including its project description, the number of stars it has received, and its contributor count. As demonstrated in the example, a contributor not only writes code but may also provide additional explanations through comments. Once the code has been compiled, if issues arise, any contributor can report an \textit{issue}, such as ``\textit{How to support function callback and Polymorphism}''. It is also possible that the same or another contributor may identify the \textit{cause} of the issue, for instance, stating that ``\textit{Class inheritance features are not complete. Operation not supported error when using super() keyword}''.

\textbf{\textit{Objectives and key Findings}}: This work aims to \textit{analyse developers’ knowledge (available GitHub projects and SO posts) to systematically and comprehensively classify the issues and root causes associated with Wasm-based applications}. To this end, we conducted an empirical investigation on 385 issues from projects hosted on GitHub and 354 issue releated questions and answers posts from SOF. %Through this analysis, we seek to provide insight into the developers' experiences and clarify the complexities involved in working with Wasm.'
The key findings of the study are structured as taxonomies of issues and their causes indicating (1) the issues in Wasm application are classified into 9 categories, of which Infrastructure and Compatibility Issues (28.16\%), Language Features and Documentation Issues (18.00\%), and Code Implementation and Build Issues (13.83\%) are the most frequently reported; and (2) the leading causes behind these issues are Syntactic and Semantic Errors (25.77\%), Configuration and Compatibility Constraints (20.1\%), and Operational Limitations (12.98\%). Primary contributions of this research are: 

%\textbf{\textit{Contributions}}: The primary contributions include (i) developing two (issue-cause) taxonomies, rooted in qualitatively analysed GitHub and SOF data in Figure \ref{fig:Taxonomy} and Figure \ref{fig:CausesTaxnomey}, as empirically derived classifications of recurring issues and their frequent causes (ii) providing publicly available data \cite{replpack}.

%%\textbf{\textit{Contributions}}: 
\begin{itemize}
    \item Mining GitHub (social coding platform) and SO (Q\&A forum) to collect and analyze practitioners' perspectives, such as code snippets, comments, scripts, queries, and responses, on predominant issues and their most frequent causes. 
    \item Taxonomic classification of issue-cause types, synthesizing available evidence (Figure \ref{fig:Taxonomy}, Figure \ref{fig:CausesTaxnomey}), to categorize, visualize, and understand the nature of issues and causes. 
    \item Providing publicly available data \cite{replpack} and outlining research implications as recommended guidelines for researching, designing, developing, and refactoring WebAssembly-based applications. The issue-cause taxonomies lay the foundation for discovering and documenting recurring solutions as patterns to address these issues (ongoing future work).
\end{itemize}

The taxonomies of issues and causes derived from our study offer a structured framework that can guide developers in diagnosing and addressing several types of problems in Wasm applications. Furthermore, these findings provide an empirical foundation for researchers to target specific areas for tool and language improvement, enhancing the overall robustness and usability of the WebAssembly ecosystem.

\section{Research Method}
\label{ResearchMethod}
\begin{figure}[!htbp]
\scriptsize
  \centering
  \includegraphics[scale=0.66]{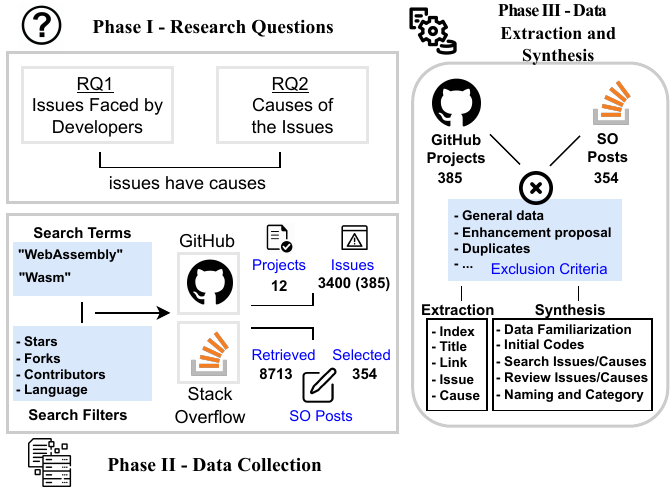}
  \caption{Overview of the research method}
  \label{fig:researchmethod}
\end{figure}

The methodology employed for this study is divided into three  phases, elaborated below and illustrated in Figure \ref{fig:researchmethod}.

%The \textit{goal} of this study is to investigate the developers discussion and SO Q\&A posts  \textit{with the purpose} to build the taxonomy of issues, and its potential causes with \textit{respect} to WebAssembly-based applications. The study has been conducted from the \textit{viewpoint} of academic researchers and seasoned software developers. The \textit{context} of the study focuses on scrutinizing developer conversations within open-source GitHub projects and Q\&A posts accessible on SO. Our study aims to provide answers to the following key research questions:

\subsection{Phase I - Research Questions (RQs)}
\label{sec:RMPhase1}

\begin{itemize}
    \item \textbf{RQ1}: \textit{What issues do developers face when working with WebAssembly applications?} The \textbf{objective} of this RQ is to systematically identify and categorize the issues faced by developers in working with WebAssembly applications. 
    \item \textbf{RQ2}:\textit{What are the causes of issues that occur in WebAssembly applications?} The \textbf{objective} of this RQ is to systematically investigate the root causes of the identified issues in WebAssembly applications.
\end{itemize}

\subsection{Phase II - Data Collection}
\label{sec:RMPhaseII}

%We collected data from two major sources. First, from GitHub open-source projects, and second, from SO. 
Data for this study was gathered from two primary platforms.

\textbf{GitHub}: We collected data from a diverse range of open-source Wasm applications hosted on GitHub (see Table \ref{tab:webassembly-projects}). To explore Wasm projects as in \ref{fig:MotivationExample}, we executed the search with the terms “\textit{WebAssembly}” and  “\textit{Wasm}” in the GitHub search bar, yielding 11,366 repository results as of March 26, 2023. We then filtered these results based on “languages” using the GitHub sidebar, resulting in 392 repositories. We discovered that several projects were aimed at developing Wasm applications but utilized other programming languages (e.g., Go, Java). To exclude such projects, we manually examined the 392 repositories, selecting only those that (i) utilized the Wasm language for over 50\% of the project, and (ii) had more than 100 closed issues. Ultimately, we identified 12 projects (see Table \ref{tab:webassembly-projects}). Our selected projects range from \textit{Runtime Environments} like Wasmer, which serves as a critical platform for executing Wasm code, to \textit{Specifications and Proposals} such as \texttt{WebAssembly/spec} and \texttt{WebAssembly/simd} that guide the platform's evolution. We also explored \textit{Toolchains and Compilers}, exemplified by projects like \texttt{WebAssembly/binaryen} and \texttt{AssemblyScript/assemblyscript}, which facilitate the development and optimization of Wasm modules. Our study also investigates \textit{Applications} related to Wasm like \texttt{torch2424/wasmboy}, a Game Boy emulator that showcases Wasm's performance capabilities, as well as \textit{Blockchain and Smart Contract systems} like \texttt{near/nearcore}, which demonstrate Wasm's versatility. Similarly, we also found \textit{UI and Frontend frameworks} for Wasm like \texttt{unoplatform/Uno.Wasm.Bootstrap}, which leverage Wasm to extend traditional web development boundaries.We extracted the developer discussions with the help of issue tracking systems of the identified 12 projects, encompassing on closed issues, in order to enhance the likelihood of discovering Wasm-related discussions. In total, we obtained 6,448 closed issues.

\textbf{Stack Overflow}: We also collected Question and Answer (Q\&A) pairs related to Wasm from SO. We initiated our data collection process by conducting an automated search on SO using two terms (i.e., “\textit{WebAssembly}”,  “\textit{Wasm}”) aligning with our GitHub search criteria. We then implemented a custom script to extract the relevant Q\&A posts retrieved from SO and store the information in data extraction sheets \cite{replpack}. This encompassed posts where the terms appeared in the post title, body of questions, and contents of answers. Initially, this search yielded 4,518 posts as of April 12, 2023.% Given the large volume of posts, manual analysis of the entire dataset was impractical. Therefore, we employed the same random sampling technique as utilized for the GitHub data source, maintaining a 95\% confidence level with a 5\% margin of error  \cite{israel1992determining}. This process resulted in a focused subset of 354 posts for our subsequent analysis.

\textbf{Random Sampling}: To conduct a comprehensive analysis of the 6,448 issue discussions from GitHub and 4,518 Q\&As from SO, we employed a random sampling formula: \[ n = \frac{N \cdot X}{X + N - 1}, \quad \text{with } X = \frac{Z^2 \cdot P \cdot (1-P)}{E^2} \]

In this formula, \( N \) is the population size, which is 6,448 and 4,518 , \( Z \) is the Z-score, which is 1.96, \( P \) is the assumed population proportion, which is 0.5, and \( E \) is the margin of error, which is 0.05.
This approach facilitated a balanced subset selection from our dataset, thereby mitigating bias and ensuring the generalizability of our findings. Additionally, it allowed for equitable comparisons among various groups within the dataset, all while optimizing resource utilization. Our selection procedure involved the random sampling of 385 issues from a pool of 6,448 issue discussions and 354 Q\&As posts from SO. These samples were drawn while maintaining a 95\% confidence level and a 5\% margin of error \cite{israel1992determining}.

\begin{table}[t]
\scriptsize
\centering
\caption{List of Identified WebAssembly Applications}
\label{tab:webassembly-projects}
\begin{tabular}{@{}lllll@{}}
\toprule
\textbf{\#} & \textbf{Project Name}  & \textbf{Closed Issues} & \textbf{Fork} & \textbf{Star} \\ \midrule
1 & Wasmerio/wasmer & 917  & 631 & 14.7K \\
2 & WebAssembly/spec & 517& 438 & 2.9K \\
3 & WebAssembly/binaryen& 693  & 640 & 6.3K \\
4 & AssemblyScript/assemblyscript& 1169 & 636 & 15.3K \\
5 & Torch2424/wasmboy & 120 & 55 & 1.3K \\
6 & WebAssembly/simd& 120  & 46 & 503 \\
7 & WebAssembly/gc& 258  & 51 & 700 \\
8 & WebAssembly/exception-handling& 89  & 33 & 115 \\
9 & Near/nearcore & 2201 & 436 & 2K \\
10 & PollRobots/scheme & 283 & 5 & 141 \\
11 & Unoplatform/Uno.Wasm.Bootstrap& 91 & 49 & 312 \\
12 & Brson/wasm-opt-rs& 13 & 6 & 22 \\ \bottomrule
\end{tabular}
\end{table}

\subsection{Phase III - Extract and Synthesize Data }
\label{sec:RMPhaseIII}
\textbf{Issues and Causes Extraction}: After selecting 12 projects, 385 developer discussions from GitHub, and 354 Q\&A posts from SO, the first and second authors manually retrieved the background information (e.g., issue label, URL) about the developer discussions and  Q\&A posts from SO. In the case of GitHub, we only selected closed issues that could potentially provide answers to our research questions. During this step, the first and second authors thoroughly analyzed each of the 385 issues and 354 Q\&A, and excluded all those that consisted of (i) general questions, opinions, feedback, and ideas; (ii) enhancement proposals; (iii) general announcements; (iv) duplicated issues or repeated questions; and (v) issues and Q\&A posts without detailed descriptions. During the data extraction, there were several discussions from GitHub and Q\&A posts from SO where the first and second authors were not able to decide whether to include them for further analysis. In such situations, the first and second authors discussed those issues with all authors to gather their opinions about inclusion or exclusion. Any disagreements about the results of the screening process were discussed among all the authors to reach a consensus.% Applying this process and criteria led to the identification of 600 posts discussing issues and 516 posts discussing causes on GitHub and SO.

\textbf{Data Extraction}: We defined a set of data items (see Table \ref{dataitems}) to answer the RQs formulated in Section \ref{sec:RMPhase1}. The first and second authors of the study conducted a pilot data extraction involving 30 GitHub discussions and 30 SO Q\&A posts, and the remaining authors evaluated the extracted data. Subsequently, the first and second authors employed a revised set of data items for formal data extraction from the selected issues. Data items (D1-D3) provide basic information, while data items (D4, D5) used to extract data to answer RQ1 and RQ2. %Spreadsheets were utilized to store the data for further synthesis.

\begin{table}
\scriptsize
\centering
\caption{Data items extracted}\label{dataitems}
%\begin{adjustbox}{width=\columnwidth,center}
\begin{tabular}{llp{5.5cm}}
\hline 
\textbf{\#} & \textbf{Data item} & \textbf{Description} \\ \hline
D1 & Index& ID of the GitHub discussion and SO post \\
D2 & Title & Title of the discussion and SO post \\
D3 & Link & Weblink of the the discussion and SO post \\
D4 & Issue & Key point(s) of the issue from discussion and posts \\
D5 & Cause & Key point(s) for the cause from discussion and posts
\\ \bottomrule
\end{tabular}
\end{table}

\textbf{Data Synthesis}: We employed the thematic analysis approach \cite{cruzes2011recommended} to analyze and classify issues and causes which is consists of five key steps: (i)\textit{Familiarizing with data}: The first and second authors thoroughly reviewed the GitHub discussion and SO pots and documented the key points related to issues and causes. (ii) \textit{Preparing initial codes}: After familiarizing, the same authors compiled an initial list of codes for the identified issues and causes (refer to the Initial Codes sheet in \cite{replpack}). (iii) \textit{Searching for the types of issues and causes}: Following the preparation of the initial codes, both the first and second authors analyzed them and categorized them into specific types of issues and causes, (iv) \textit{Reviewing types of issues and causes}: All authors collaboratively reviewed and refined the coding results, organizing them under the relevant types of issues and causes. During this process, we engaged in discussions, separating, merging, or discarding several issues and causes. (v) \textit{Defining and naming categories}: We precisely defined and further refined all types of issues and causes by creating clear subcategories and categories. By following these steps, we established three levels of categories for effectively managing the identified issues and causes for Wasm applications.
\section{Results -- Issues and Causes in Wasm }
\label{sec:Results}
This section presents the study results, addressing the two RQs outlined in Section \ref{sec:RMPhase1}. The results are organized into categories, subcategories, and types. Categories are presented in \textbf{boldface}, subcategories in \textit{italic}, and types in \textsc{small capitals}. At the end of each section, based on the study results, a ‘Takeaways’ box provides the key messages for Wasm researchers and practitioners.

\subsection{Types of Issues (RQ1)}
\label{sec:results_RQ1}
\begin{figure*}[!htbp]
\scriptsize
  \centering
  \includegraphics[scale=0.33]{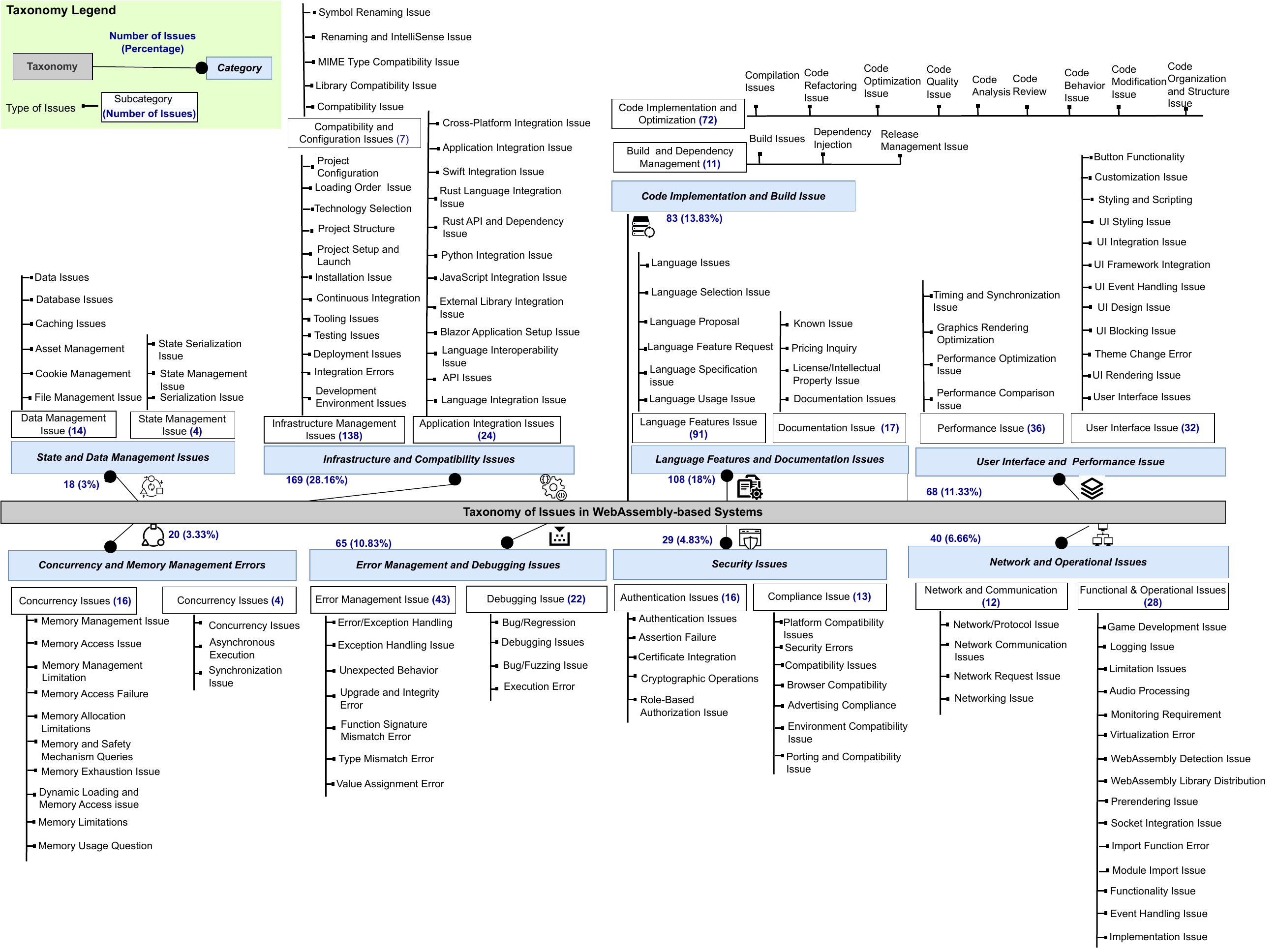}
        \caption{A Taxonomy of Issues in Wasm Applications}
    \label{fig:Taxonomy}
\end{figure*}

The taxonomy of Wasm application issues is shown in Figure \ref{fig:Taxonomy}. Developed from analyzing GitHub developer discussions and SO Q\&As, it categorizes 739 issues into 9 main categories with 19 subcategories, totaling 120 types. Descriptions of each category are provided below, with detailed data in our replication package \cite{replpack}.

%The taxonomy of issues related to Wasm applications is illustrated in Figure \ref{fig:Taxonomy}. This classification was developed by analyzing developer discussions and SO posts, encompassing total of 739 instances of issues taxonomically classified as 9 categories having 19 subcategories that has 130 types. Each category in the taxonomy is briefly described below, and more detailed information is available in the study's replication package \cite{replpack}.

\textbf{1. Infrastructure and Compatibility Issues}: This category broadly covers issues related to system architecture, integration of various components, and compatibility issues in Wasm applications. It is composed of three subcategories: \textit{Infrastructure Management Issues}, which collect concerns in setting up and maintaining the infrastructure necessary for Wasm; \textit{Application Integration Issues}, which gather problems in the integration of Wasm with various programming languages and platforms; and \textit{Compatibility and Configuration Issues}, which amass issues related to the compatibility of Wasm across different systems. In total, there are 169 issues, constituting 28.16\% of all identified issues.

Examples of issues within these subcategories include \textsc{testing issues}, \textsc{tooling issues}, and \textsc{integration issues}, which are related to compromising the reliability and efficiency of the infrastructure. Additionally, \textsc{compatibility issues} and \textsc{symbol renaming issues}, are crucial to ensure that Wasm modules operate correctly across different environments. Tackling these issues is critical for the robust deployment and functioning of Wasm applications.

\textbf{2. Language Features and Documentation Issues}: This category encompasses concerns related to the features of the programming languages that are available in Wasm, along with issues pertaining to the associated documentation. It consists of two subcategories: \textit{Language Feature Issues}, which collects challenges related to language use, specifications, and the introduction of new features, and \textit{Documentation Issues}, which aggregates problems involving existing documentation, licensing, intellectual property rights, and queries regarding pricing. Altogether, there are 108 issues noted, which constitute 18.00\% of all issues identified.

Within these subcategories, examples of issues such as \textsc{language usage issues}, \textsc{language specification issues}, and \textsc{language feature requests} are significant, as they directly influence the efficacy with which developers can leverage Wasm. In parallel, \textsc{documentation issues} and \textsc{license/intellectual property issues} underscore the importance of having clear, accessible, and legally robust support materials to facilitate the adoption and effective use of Wasm. Addressing these issues is essential for fostering a comprehensive understanding and application of Wasm within the developer community.

\begin{comment}
\textbf{2. Language Features and Documentation Issues} (18.00\%): This category focuses on issues related to the features and usage of the programming language(s) used in Wasm applications, as well as relevant documentation, licensing, and pricing concerns.
\begin{itemize}
\item \textit{Language Features Issue}: This subcategory includes issues related to the language features and how they are used in Wasm projects. In this subcategory, most of issues are related to  \textsc{language feature request}, \textsc{language proposal}, and \textsc{language specification} issues.
\item \textit{Documentation and Licensing Issues}: This subcategory collect issues related to project documentation, licensing terms, and pricing inquiries. Issue types include \textsc{documentation issues}, \textsc{license/intellectual property issue}, \textsc{pricing inquiry}, and \textsc{known issue}. 
\end{itemize}
    \end{comment}

\textbf{3. Code Implementation and Build Issue}: This category is concerned with challenges encountered during the coding and build phases in software development, which are especially pertinent for Wasm given its need for compilation. It includes two main subcategories: \textit{Code Implementation Issues}, representing the range of problems that can arise during the actual coding process, and \textit{Build Issues}, that pertain to complications that occur during the software build process, such as dependency management. This category has a total of 83 issues, representing 13.83\% of all the issues identified.

Issues within these subcategories, such as \textsc{code quality issues}, \textsc{code analysis issues}, and \textsc{code review and feedback issues}, are crucial as they directly impact the efficacy and maintainability of Wasm modules. The compilation process and associated challenges, including \textsc{build issues} and \textsc{dependency management}, are important to address because they affect the performance, reliability, and the smooth deployment of Wasm applications, thus influencing their stability and the cycle of updates.

\begin{comment}
\textbf{3. Code Implementation and Build Issues} (13.83\%): This category collect and categorized  a range of issues related to the development and construction of software code, as well as the management of dependencies and builds.

\begin{itemize}
    \item \textbf{Code Implementation and Optimization Issues}: This subcategory capture issues primarily dealing with the behavior, quality, and organization of code. It includes issues like \textsc{code behavior issue}, \textsc{code modification issue}, and \textsc{code optimization issue}.
   \item \textbf{Build and Dependency Management Issues}: This subcategory deals with issues related to the build process and dependency management, such as \textsc{build issue}, \textsc{dependency injection}, and \textsc{release management issue}.
\end{itemize}
\end{comment}

\textbf{4. User Interface and Performance Issue}: This category encapsulates concerns with the user-facing aspects and the efficiency of Wasm applications. It is divided into \textit{User Interface Issues}, which pertains to the design and interactivity components, including elements like button functionality and UI customization, and \textit{Performance Issue}, which deals with the speed and responsiveness of the application, including performance optimization and graphics rendering. In total, this category includes 68 issues, constituting 11.33\% of all issues identified.

Within these subcategories, specific challenges like \textsc{ui rendering issues} and \textsc{ui design issues} are crucial as they directly influence user engagement and satisfaction. Performance concerns, such as \textsc{performance optimization issues} and \textsc{timing and synchronization issues}, are fundamental to the functionality of Wasm applications, affecting their operational capability and the user experience. Addressing these issues is critical to enhancing both the interface and the performance of Wasm applications for end users.
\begin{comment}

\textbf{4. User Interface and  Performance Issues} (68/600, 11.33\%): This category collect issues related to user interface (UI) design, customization options, and performance bottlenecks in Wasm applications. Below is a brief description of each subcategory.

\begin{itemize}
\item \textit{Performance and Optimization Issues}: This subcategory captures issues related to system performance and optimization. Types of issues in this subcategory include \textsc{performance comparison issue}, \textsc{performance optimization issue}, \textsc{graphics rendering optimization}, and \textsc{timing and synchronization issue}.
\item \textit{User Interface  Issue}: This subcategory collect issues related to user experience (UX) and includes issues such as \textsc{theme change error}, \textsc{ui blocking issue}, and \textsc{ui design issue}.
\end{itemize}
\end{comment}

\textbf{5. Error Management and Debugging Issues}: This category addresses the crucial aspects of identifying, handling, and resolving errors in Wasm applications. It is categorized into \textit{Debugging Issues}, which includes problems like bug regressions, debugging intricacies, and bug fuzzing, and \textit{Error Management Issue}, which covers the systematic approach to error and exception handling, and issues arising from unexpected behavior or integrity errors. There are 65 issues in total, accounting for 10.83\% of all issues documented.

Specifically, within these subcategories, challenges such as \textsc{execution errors}, \textsc{error/exception handling}, and \textsc{unexpected behavior} are pivotal, as they impact the stability and reliability of Wasm applications. Issues like \textsc{function signature mismatch error}, \textsc{type mismatch error}, and \textsc{value assignment error} underscore the complexities of ensuring accurate execution and data integrity. Effective management and resolution of these issues are essential for the development of robust, error-resistant Wasm applications.

\begin{comment}
    
\textbf{5. Error Management and Debugging Issues} (10.83\%): This category group the issues related to error management and the debugging process in software development. We identified and classified 11 types of issues in 2 subcategories (see Figure \ref{fig:Taxonomy} and the Issue Taxonomy sheet in \cite{replpack}). Each of them is briefly described below.

\begin{itemize}
\item \textit{Error Management Issue}: This subcategory deals with how errors and exceptions are managed in the code. It consist of 7 types in which leading three are \textsc{error\/exception handling}, \textsc{unexpected behavior}, and \textsc{upgrade and integrity error}.
\item \textit{Debugging Issue}: This subcategory focuses on issues related to debugging Wasm projects. Specific types of issues include \textsc{bug\/fuzzing issue}, \textsc{bug\/regression}, and \textsc{execution error}.
\end{itemize}
\end{comment}

\textbf{6. Network and Operational Issues}: This category pertains to challenges associated with networking and the day-to-day operational aspects of Wasm applications. It is subdivided into \textit{Functional \& Operational Issues}, which comprise problems affecting application functionality and operations such as event handling, module imports, and system integrations, and \textit{Network and Communication Issues}, which deal with data transmission, protocol adherence, and network requests. There are 40 documented issues in total, which account for 6.66\% of all issues reported.

In these subcategories, specific concerns like \textsc{functionality issues}, \textsc{prerendering issues}, and \textsc{socket integration issues} are significant as they directly influence the operational effectiveness of Wasm applications. Network-related issues such as \textsc{network/protocol issues} and \textsc{network communication issues} are critical for maintaining robust communication channels within and across Wasm applications. %Efficiently addressing these network and operational issues is crucial for ensuring the seamless functioning and integration of Wasm in various networked environments.

\textbf{7. Security Issues}: This category encompasses the various aspects of security within Wasm applications. It includes \textit{Authentication Issues}, which cover problems related to user verification, assertion checks, certificate integration, cryptographic operations, and role-based authorization. Another critical area is \textit{Compliance Issues}, comprising 13 issues related to adherence to platform standards, browser compatibility, and environmental regulations, as well as the challenges in porting applications while maintaining compliance. There are 29 documented issues in total, which account for 4.83\% of all issues reported.

Specific challenges within these subcategories, such as \textsc{authentication issues} and \textsc{cryptographic operations}, are fundamental to the secure operation of Wasm applications. Compliance concerns, including \textsc{platform compatibility issues} and \textsc{advertising compliance}, are crucial for the applications to operate within the legal and technical frameworks of various environments. Effectively managing these security and compliance issues is paramount for the integrity and reliability of Wasm applications.

\textbf{8. Concurrency and Memory Management Errors}: This category addresses critical issues related to the simultaneous operation of multiple processes and the efficient management of memory in Wasm applications. It accounts for 20 issues in total, comprising 3.33\% of all recorded problems. Within this category, there are \textit{Concurrency Issues}, which include challenges like managing asynchronous execution and synchronization, and \textit{Memory Management Errors}, which involve a variety of concerns ranging from memory access and allocation issues to questions about memory usage and the limitations inherent in dynamic loading. There are 20 issues in total, comprising 3.33\% of all recorded problems.

\textbf{9. State and Data Management}: This category is concerned with issues related to maintaining the state of applications and the management of data within Wasm applications, accounting for 18 issues and representing 3\% of all issues. It is divided into \textit{State Management Issues}, which includes problems like state serialization and general state management concerns, and \textit{Data Management Issues}, which covers a broader range of data-related challenges such as database integration, caching strategies, asset management, cookie handling, and file management issues.

Within these subcategories, issues such as \textsc{state serialization issues} and \textsc{serialization issues} are critical because they affect how application state is maintained and restored, which is vital for the user experience. On the data management side, issues like \textsc{database issues}, \textsc{caching issues}, and \textsc{file management issues} are essential for the efficient operation and scalability of Wasm applications. %Addressing these state and data management issues is key to ensuring that Wasm applications function reliably and manage resources effectively.
\begin{comment}

\textbf{9. State and Data Management Issues} (3.00\%): This category encompasses issues related to the management of state and data within Wasm applications. It consists of two main subcategories: \textit{State Management and  Serialization} and \textit{Data and Asset Management}.% Each of these subcategories, along with their specific types of issues, is further reported in Figure \ref{fig:Taxonomy} and the Issue Taxonomy sheet in \cite{replpack}. %Each subcategory is briefly described below.
\begin{itemize}
\item \textit{Data and Asset Management Issues}: This subcategory collect issues concerning the management of data and digital assets. Types of issues in this category include \textsc{Asset Management}, \textsc{Data Issues}, \textsc{database issues}, \textsc{caching issues}, \textsc{cookie management}, and \textsc{file management issue}. %A representative issue could be "\textit{Database connection fails during high-traffic scenarios, \#XX}".
\item \textit{State Management and Serialization Issues} (4, 0.33\%): This subcategory group issues related to the management and serialization of application state. It includes specific types of issues such as \textsc{state management issue}, \textsc{serialization issue}, and \textsc{state serialization issue}. 
\end{itemize}
\end{comment}

\begin{takeawaybox}{Takeaways}
\scriptsize
 \blackcircle{1} \textbf{Infrastructure and Compatibility}: Leading issues include system architecture and API complexities, affecting Wasm's seamless integration with existing systems.\\
\blackcircle{2}\textbf{Operational Issues}: Networking and communication issues notably impact the stability and reliability of Wasm applications.\\
 \blackcircle{3} \textbf{Code and Build Issues}: Implementation, optimization, and dependency management are key areas needing attention for quality Wasm application development.
\end{takeawaybox}

\subsection{Causes of Wasm Issues (RQ2)}
\label{sec:causes_RQ2}
\begin{figure*}[!htbp]
 %   \flushleft
 \scriptsize
 \centering
      \includegraphics[scale=0.25]{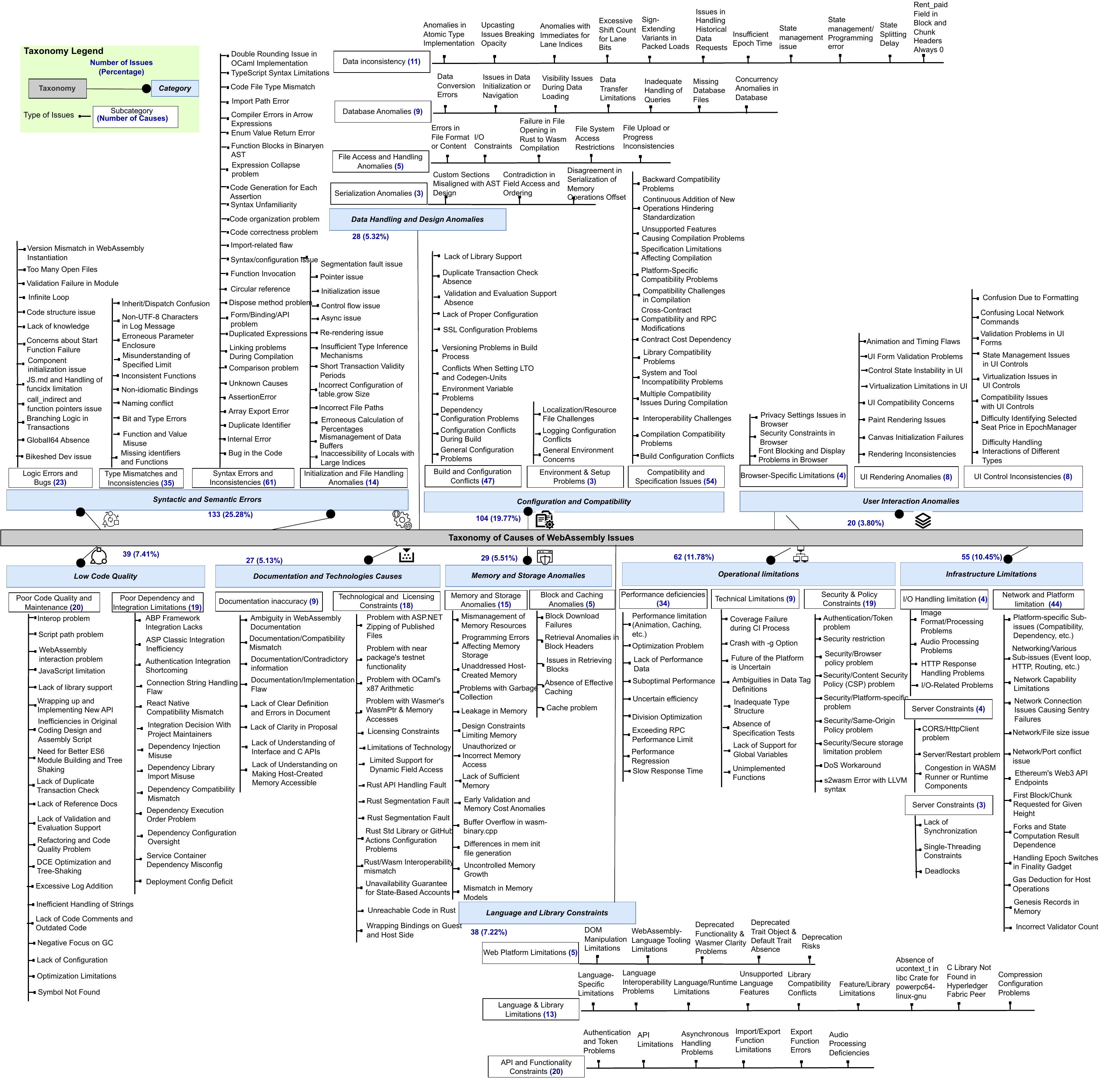}  
        \caption{A Taxonomy of Causes of Issues in Wasm Applications}
    \label{fig:CausesTaxnomey}
\end{figure*}

The taxonomy of causes of Wasm issues is detailed in Figure \ref{fig:CausesTaxnomey}. The cause taxonomy is based on data mined from developer discussions on GitHub and SO. It is important to note that not all discussions on these platforms provide cause information. Therefore, we identified only 516 cause instances from both sources.  This analysis identified 278 cause types, categorized into 10 main categories and 29 subcategories. Detailed information is available in the dataset \cite{replpack}. %The results show that \textbf{Syntactic and Semantic Errors}, \textbf{Configuration and Compatibility}, and \textbf{Operational limitations} are the top-level categories of causes. Each cause category is briefly reported below.

\textbf{1. Syntactic and Semantic Errors}: This cause category encompasses causes that originate from syntactic and semantic inconsistencies within Wasm code, often leading to compilation and runtime issues , or unexpected behavior. It includes a total of 133 reported causes. Subcategories within this category are \textit{Syntax Errors and Inconsistencies}, \textit{Initialization and File Handling Anomalies}, \textit{Type Mismatches and Inconsistencies}, and \textit{Logic Errors and Bugs}. Some of the leading key types of causes such as \textsc{bug in the code}, \textsc{internal error}, and \textsc{syntax unfamiliarity} within \textit{Syntax Errors and Inconsistencies} are critical as they directly affect the correct interpretation and execution of Wasm code. In \textit{Initialization and File Handling Anomalies}, causes like \textsc{mismanagement of data buffers} and \textsc{segmentation fault issues} are significant, as they can lead to crashes and unpredictable behavior. \textit{Type Mismatches and Inconsistencies} include crucial causes such as \textsc{missing identifiers and functions}, which can prevent code from compiling or running correctly, while \textit{Logic Errors and Bugs}, with causes like \textsc{branching logic in transactions} and \textsc{component initialization issues}, can result in flawed application logic and runtime errors.

\textbf{2. Configuration and Compatibility}: This cause category is central to issues that arise from the setup and interoperability of Wasm systems, featuring 104 reported causes. Within this category, we have three subcategories: \textit{Compatibility and Specification Issues}, \textit{Build and Configuration Conflicts}, and \textit{Environment \& Setup Issues}.

Notable causes within \textit{Compatibility and Specification Issues} include \textsc{interoperability challenges} and \textsc{library compatibility issues}, which are critical for ensuring that Wasm modules work across different platforms and with various libraries. In \textit{Build and Configuration Conflicts}, causes such as \textsc{environment variable issues} and \textsc{ssl configuration issues} can lead to significant deployment problems. Moreover, \textit{Environment \& Setup Issues} like \textsc{logging configuration conflicts} affect the operational aspect of Wasm applications, highlighting the need for meticulous configuration management. Addressing these configuration and compatibility causes is essential to prevent disruptions in Wasm application development and deployment.

\textbf{3. Operational Limitations}: This cause category deals with constraints that impact the functionality and security of Wasm applications during their operation, totaling 62 reported causes. Within this category, we have two subcategories: \textit{Performance deficiencies}, \textit{Technical Limitations}, and \textit{Security Constraints}.
Key types of causes within \textit{Performance deficiencies}, such as \textsc{performance limitation} and \textsc{performance regression}, are critical as they can significantly degrade user experience and application responsiveness. Within \textit{Technical Limitations}, causes like \textsc{lack of support for global variables} and \textsc{absence of specification tests} pose challenges for developers by restricting the functionality and verifiability of Wasm modules. \textit{Security Constraints} involve issues like \textsc{authentication/token issues} and \textsc{security/browser policy issues}, which are essential for maintaining the integrity and trustworthiness of applications. Effectively addressing these operational limitations is crucial for the advancement and secure deployment of Wasm applications.

\textbf{4. Infrastructure Limitations}: This cause category encompasses foundational concerns that impact the operation of Wasm applications. It includes two subcategories: \textit{Network and Platform limitation}, \textit{I/O Handling limitation} and \textit{Synchronization Limitation}.

Significant causes within \textit{Network and Platform limitation} involve challenges like \textsc{network connection issues} and \textsc{platform-specific compatibility issues}, which can severely restrict an application's functional scope and connectivity. \textit{I/O Handling limitation} is vital for the application’s interface with the user and the system, where issues such as \textsc{HTTP response handling} and \textsc{CORS issues} are key operational concerns. In the realm of \textit{Synchronization Limitation}, causes such as \textsc{single-threading constraints} and \textsc{deadlocks} highlight the complexities of managing concurrent operations in Wasm. Effectively tackling these infrastructure limitations is essential for the seamless operation and scalability of Wasm applications.

\textbf{5. Low Code Quality}: Gathers the causes related to inadequately maintained and poor-quality codebases in JavaScript and Wasm interactions, along with a deficiency in essential supportive elements such as libraries and documentation. It includes \textit{Poor Code Quality and Maintenance} and \textit{Poor Dependency and Integration Limitations} subcategories.

Key causes such as \textsc{script path issues} and \textsc{inefficiencies in original coding design} from the \textit{Poor Code Quality and Maintenance} subcategory are pivotal, as they can directly impact the functionality and extendibility of the code. Additionally, \textit{Poor Dependency and Integration Limitations} present significant causes like \textsc{authentication integration shortcomings} and \textsc{service container dependency misconfigurations}, which can complicate the integration process and affect the stability of the application. Addressing these causes is essential for the development of high-quality, maintainable Wasm applications that are well-integrated within their respective ecosystems.

\begin{comment}
    
\textbf{5. Low Code Quality} (7.55\%): This category encapsulates causes related to the inferior quality of code, reflective of poor coding practices and maintenance, which can lead to various challenges such as JavaScript/Wasm interaction issues and limitations, and inefficiencies in original coding design.  We identified and classified 35 types of Low code quality in 2 subcategories (see Figure \ref{fig:CausesTaxnomey} and the Cause Taxonomy sheet in \cite{replpack}). Each of them is briefly described below.

\begin{itemize}
\item \textit{Poor Code Quality and Maintenance}: This subcategory gathers the causes related to inadequately maintained and poor-quality codebases in JavaScript and Wasm interactions, along with a deficiency in essential supportive elements such as libraries and documentation. We have identified 19 distinct causes within this subcategory, with the prominent three are \textsc{javascript/interop issue}, \textsc{javascript/script path issue}, and \textsc{javascript-Wasm interaction issue}
\item \textit{Poor Dependency and Integration Limitations}: This subcategory encompasses causes related to difficulties in integration and dependencies, reflecting compatibility issues. These issues illuminate the complexities in integrating diverse frameworks and managing dependencies in Wasm applications. This subcategory consists of 14 distinct causes, with the prominent three are \textsc{integration issue}, \textsc{integration/abp framework issue}, and \textsc{integration/asp classic issue}.
\end{itemize}
\end{comment}

\textbf{6. Language and Library Constraints}:Consolidates the causes related to the limitations within programming languages and their associated libraries in the context of Wasm. This category combines causes into three subcategories \textit{API and Functionality Constraints}, \textit{Language \& Library Limitations}, and \textit{Web Platform Limitations} subcategories.

For example, \textit{API and Functionality Constraints} involve critical causes such as \textsc{API limitations} that can significantly hamper the integration and operational capabilities of Wasm modules. \textit{Language \& Library Limitations} highlight causes like \textsc{language interoperability issues}, which affect the seamless integration of Wasm with other programming environments. Furthermore, \textit{Web Platform Limitations} bring attention to causes such as \textsc{tooling limitations}, emphasizing the need for up-to-date and compatible tools to support the evolving landscape of Wasm. Navigating these causes is key to enhancing Wasm's adaptability and ensuring its effective deployment across various platforms.

\textbf{7. Documentation and Technologies Causes}: This category identifies causes related to informational discrepancies and technical limitations that affect the use and development of Wasm. This category combines causes into two subcategories: \textit{Documentation Inaccuracy} and \textit{Technological and Licensing Constraints}.

For instance, \textit{Documentation inaccuracy} covers causes such as \textsc{lack of clear definition and errors in document}, which can create barriers to correctly implementing and leveraging Wasm's functionalities. \textit{Technological and Licensing Constraints} highlight issues like \textsc{rust segmentation fault} and \textsc{rust/wasm interoperability mismatch}, pinpointing the technical hurdles that can arise due to language-specific features or the integration of different technologies. These constraints underscore the importance of accurate documentation and adaptable technology solutions to support the evolving needs of Wasm applications and their users. Addressing these causes is essential to foster a clear understanding and effective utilization of Wasm across various domains.

\begin{comment}
    
\textbf{7. Documentation and Technologies Causes} (5.23\%):
This category captures causes related to flaws in documentation and constraints in technologies and licensing within Wasm development, leading to incorrect implementations and limitations in the optimal use of technologies. Each of the derived subcategory is briefly described below.

\begin{itemize}
\item \textit{Technological and Licensing Constraints}: The causes in this subcategory are related to technological limitations and licensing constraints. They reflect challenges and restrictions associated with the use of various tools, packages, and licensing in Wasm, which can impact the reliability and efficiency of the developed applications. This subcategory includes 17 distinct causes, with the top two being: \textsc{issue with asp.net zipping of published files} and \textsc{issue with near package's testnet functionality}.

\item \textit{Documentation Inaccuracy}: This subcategory addresses problems caused by inaccurate or unclear documentation. The challenges here arise from a lack of clear definitions and contradictory information in the Wasm documentation. Within this subcategory, eight distinct causes have been identified, among which the top two are \textsc{ambiguity in Wasm documentation} and \textsc{documentation/compatibility issue}.
\end{itemize}
\end{comment}

\textbf{8. Data Handling and Design Anomalies}: This category captures causes concerning the integrity and structure of data within Wasm applications. It category combines causes into three subcategories: \textit{Data inconsistency}, \textit{Database Anomalies}, \textit{File Access and Handling Anomalies}, and \textit{Serialization Anomalies}.

For example, causes such as \textit{anomalies in atomic type implementation} and \textit{state management issues} within the \textit{Data inconsistency} subcategory can directly affect the accuracy and reliability of data processes. Causes in \textit{Database Anomalies}, like \textit{data conversion errors} and \textit{missing database files}, are pivotal as they influence the robustness of database operations. Within \textit{File Access and Handling Anomalies}, causes such as \textit{I/O constraints} and \textit{file system access restrictions} can severely limit application functionality. Causes in \textit{Serialization Anomalies}, including \textit{discrepancies in serialization of memory operations offset}, can lead to data integrity concerns. %Identifying and addressing these causes is essential to enhance data management practices and ensure that Wasm applications handle data effectively and maintain their intended functionality.

\textbf{9. Memory and Storage Anomalies}: This category collects causes associated with the mismanagement and technical challenges of memory and storage within Wasm applications. It combines causes into two subcategories \textit{Memory and Storage Anomalies} and \textit{Block and Caching Anomalies}.

Causes like \textit{mismanagement of memory resources} and \textit{issues with garbage collection} within the \textit{Memory and Storage Anomalies} subcategory are critical as they directly influence the application's stability and resource optimization. In the \textit{Block and Caching Anomalies} subcategory, causes such as \textit{block download failures} and \textit{cache problems} highlight the importance of reliable data storage and efficient retrieval mechanisms. Addressing these memory and storage causes is fundamental to ensuring that Wasm applications maintain their integrity and provide a responsive user experience.
\begin{comment}

\textbf{9. Memory and Storage Anomalies} (3.87\%):
This category dives into causes related to anomalies in memory and storage within Wasm applications. It addresses issues arising from the mismanagement of memory resources, unauthorized or incorrect memory access, and disparities in memory models which could potentially lead to suboptimal performance and system instabilities.
\begin{itemize}
\item \textit{Memory and Storage Anomalies}: This subcategory groups causes related to the mismanagement of memory resources and programming errors affecting memory storage. It explores the intricacies involved in memory management, detailing issues such as memory leakage and insufficient memory. Within this subcategory, twelve distinct causes are identified, with the top three being \textsc{mismanagement of memory resources}, \textsc{programming errors affecting memory storage}, and \textsc{unaddressed host-created memory}.
\item \textit{Block and Caching Anomalies}: This subcategory is centered around causes resulting from anomalies in block retrieval and caching. Five unique cause
s have been identified within this category, with the top three causes being \textsc{block download failures}, \textsc{retrieval anomalies in block headers}, and \textsc{issues in retrieving blocks}.
\end{itemize}
\end{comment}

\textbf{10. User Interaction Anomalies}: This category encompasses causes that negatively impact the user's ability to interact with Wasm applications effectively. It includes \textit{UI Control Inconsistencies}, \textit{UI Rendering Anomalies}, and \textit{Browser-Specific Limitations} subcategories. For instance, within \textit{UI Control Inconsistencies}, causes such as \textit{compatibility issues with UI controls} and \textit{confusion due to formatting} can disrupt the user’s navigation and interaction with the application. \textit{UI Rendering Anomalies} highlight causes like \textit{animation and timing flaws}, which are crucial for a seamless and intuitive user interface. \textit{Browser-Specific Limitations} bring to light causes such as \textit{security constraints in browser}, which can limit functionality and affect the overall accessibility of Wasm applications. %Addressing these user interaction anomalies is crucial for providing a smooth and engaging experience for users across different platforms and browsers.

\begin{takeawaybox}{Takeaways}
\scriptsize
 \blackcircle{4} \textbf{Diversity in Language Compilation}: The diversity of source languages compilable to Wasm causes inconsistencies and errors in resultant applications.\\
 \blackcircle{5} \textbf{Security Vulnerabilities due to Wasm’s Structure and Execution Model}: Wasm’s structure and execution model are inherent causes of new security vulnerabilities in applications.\\
 \blackcircle{6} \textbf{Complexities in Optimizing Compiled Code}: The inherent complexities in optimizing Wasm code cause performance bottlenecks affecting user experience.
\end{takeawaybox}

\section{Discussion and Implications}
\label{sec:Discussion}
This section presents the discussion on the key takeaways along with implications for researchers and practitioners based on the study results. Section 4.1 outlines the potential implications associated with WebAssembly issues, while section 4.2 discusses into the various WebAssembly causes.
\subsection{Wasm Issues}
\label{WebAssemly issues}
\blackcircle{1} \textbf{Infrastructure and Compatibility Issues}: Among the key insights gained from mining GitHub and SO discussions is the recurring theme of infrastructure and compatibility issues with Wasm. The consistent mention of these challenges among developers suggests that integrating Wasm into existing systems remains a significant obstacle. This aligns with existing studies (e.g., \cite{li2023warnings}) highlight the difficulty in adopting new technologies due to system architecture complexities \cite{li2023warnings} and API-related issues \cite{li2018cid}. The online developer discussions illuminate a gap between academic understanding and real-world practice. Although the academic literature may describe the theoretical benefits of Wasm \cite{eleskovic2020closer}, the actual integration into existing architectures proves to be a complex issue not fully addressed \cite{bosshard2020use}. \colorbox{lightgray}{\textit{Implications}}: The persistent nature of these issues highlights the need for academic research that bridges theory and practice, focusing on creating more robust integration methods or frameworks for Wasm. Practitioners can benefit from more actionable guidance, possibly in the form of best practice documents.

\blackcircle{2} \textbf{Operational Issues}: A critical aspect uncovered pertains to operational challenges in networking and communication, as highlighted in discussions across platforms like GitHub and SO. Developers have raised concerns about the reliability of these aspects in Wasm \cite{haas2017bringing}, echoing literature that points to new technologies often grappling with underdeveloped networking protocols (e.g.,\cite{ray2023overview}). There is a discernible lack of literature focusing specifically on Wasm's operational capabilities in these areas. Insights from real-world scenarios, as seen in the aforementioned platforms, are crucial in bridging this knowledge gap \cite{vsipek2021next}. \colorbox{lightgray}{\textit{Implications}}: This presents a fertile ground for researchers to delve into Wasm's networking functionalities and suggest improvements. Similarly, practitioners are advised to examine the operational elements of Wasm meticulously, potentially utilizing external libraries or modules as interim solutions.

\blackcircle{3} \textbf{Code Implementation and Optimization Issues}: Data mined from developer conversations also highlight challenges in code implementation and optimization, including build and dependency management. The online discussions complement the literature (e.g., \cite{wang2021empowering, haas2017bringing}) which often talks about the lack of mature toolsets for new technologies, emphasizing that Wasm development is still in its infancy stage \cite{niessen2020insights, vsipek2021next}. Although academic discourse may emphasize the computational efficiency of Wasm, it seems to overlook the practical aspects of code implementation and optimization, an area clearly fraught with challenges according to GitHub and SO data. \colorbox{lightgray}{\textit{Implications}}: Researchers could aim to develop better tools for Wasm development, possibly in collaboration with industry stakeholders, to address the implementation challenges. Practitioners could consider incorporating emerging best practices and tools as they become available, staying up-to-date through both academic and community channels.

\subsection{Causes of Wasm Issues}
\label{CausesIssues}
%The study summarized three key findings related to causes of issues in Wasm-based applications.%: the diversity in language compilation is creating significant problems, the unique structure and execution model of Wasm are causing security vulnerabilities, and complexities in optimizing the compiled code are leading to performance bottlenecks.

\blackcircle{4} \textbf{Diversity in Language Compilation}: We find the diversity in language compilation to Wasm is causing significant discrepancies and anomalies in the resultant applications, creating inconsistencies in application behavior and functionality. This issue is consistent with existing studies (e.g., \cite{infoworld2023}) and practitioner perspective (e.g., \cite{plainenglish2023, cloudflare2023, infoworld2023}), highlighting the challenges and irregularities arising due to compiling a variety of languages like C, C++, and Rust to Wasm. It reaffirms the prevailing knowledge base, emphasizing the problems in maintaining consistency during compilation processes. The diverse origin of source languages necessitates a more universal and standardized compilation strategy to prevent the resultant inconsistencies and anomalies in Wasm-based applications. \colorbox{lightgray}{\textit{Implications}}: There is a need to develop more comprehensive and robust compilation methods to accommodate the diversity in source languages. Developers should be aware of the complications arising from language diversity and consider the compatibility of origin languages with Wasm during the development phase.

\blackcircle{5} \textbf{Security Vulnerabilities due to Wasm's Structure and Execution Model}: Our research identifies that the unique structure and execution model of Wasm are introducing new security vulnerabilities and expanding the application's attack surface. This finding align with some of the earlier studies (e.g., \cite{stievenart2021security, lehmann2020everything}) that depicted Wasm as a more secure alternative to JavaScript. The findings also reveals potential gaps in our understanding of Wasm's security model and necessitates further exploration into its unique vulnerabilities. The alignment between our results and previous studies highlights the evolving and dynamic nature of Wasm, suggesting continuous emergence and evolution of potential security threats and vulnerabilities. \colorbox{lightgray}{\textit{Implications}}:
This contradiction prompts a deeper examination of Wasm’s security framework, urging further exploration and research into its vulnerabilities and mitigation strategies. Developers need to implement rigorous security protocols and continuously monitor and update the security features of applications to mitigate the risks associated with the unique vulnerabilities of Wasm. 

\blackcircle{6} \textbf{Complexities in Optimizing Compiled Code}: The research indicates that the complexities involved in optimizing the compiled code are creating substantial performance bottlenecks, affecting the user experience and application response times adversely. The findings align well with existing literature (e.g., \cite{cabrera2020superoptimization, jangda2019not, wang2021empowering}), emphasizing the critical need to address these performance bottlenecks by developing advanced optimization techniques to improve the efficiency and response time of Wasm applications. \colorbox{lightgray}{\textit{Implications}}: The recurring issues related to performance bottlenecks in our findings indicate the need of optimization techniques and methodologies to enhance the user experience and application efficiency. Developers and IT professionals should prioritize resolving these performance bottlenecks by exploring and implementing new optimization solutions and techniques to enhance application performance and user experience.

\section{Related Work}
\label{RelatedWork}
This section overviews the most relevant existing research, classifying and analyzing empirically-based studies focused on (i) bugs and security issues along with (ii) performance challenges in Wasm applications. A conclusive summary highlights the scope and contributions of the proposed research in the context of related work.

\subsection{Bugs and Security Issues in Wasm} 
Bugs in Wasm applications are among the prevailing challenges including issues that relate to not the bugs, errors, and security risks during application compilation  \cite{chen2022wasai, quan2019evulhunter, zhou2023wasmod}. Specifically, Romano et al. \cite{9678776} conducted an empirical study to analyze 1,054 bugs in Wasm compilers The study investigated `lifecycle', `impact', and `sizes' of bug-inducing inputs and bug fixes and highlighted the need for further research on principles and practices to debug Wasm applications. %In line with the findings of \cite{9678776}, other studies (e.g., [38], [39]) have also focused on runtime bugs that can be the root cause of security issues. For example, the study [38] experimentally analyzed four popular WebAssembly runtimes (V8, SpiderMonkey, Wasmer, and Wasmtime) to report issues relating to bug detection, localization, debugging, and repair of Wasm applications. 
Security-critical issues in Wasm have gained significant attention of researchers with web application development for blockchain solutions \cite{10123536, yixuan2023characterizing}.  Lehmann et al. \cite{lehmann2020everything} examined security vulnerabilities to analyze the extent vulnerabilities are exploitable in WebAssembly binaries, and how this compares to native code in Wasm and proposed solutions. Similar studies such as \cite{chen2022wasai, quan2019evulhunter, zhou2023wasmod} address the security of Wasm-based smart contracts for blockchain systems. Compared to conventional Ethereum smart contracts, Wasm smart contracts have shown growing popularity for web-based blockchains, however, they suffer from various attacks exploiting their vulnerabilities \cite{quan2019evulhunter}.% Chen et al \cite{chen2022wasai} have designed and developed WASAI, a new  fuzzer for uncovering vulnerabilities in Wasm smart contracts. 

\subsection{Performance Issues}
Performance issues in Wasm applications can jeopardise time-critical transactions and user experience in web systems.  The research by Jangda et al. \cite{jangda2019not} empirically compares native and Wasm code to identify the bottlenecks that slow down application execution. A similar study by Yan et al. \cite{yan2021understanding} compared Wasm and JavaScript performance to guide developers in identifying optimization opportunities in web development.  Furthermore, Andre et al. \cite{andre2022developers} investigate Wasm-related discussions on Stack Overflow, revealing security concerns and frequent requests for bug-fixing corresponding to the performance of Wasm-based web applications. 

\textbf{\textit{Conclusive Summary}}: Based on the review above, we conclude that the proposed research is closely aligned and complements the existing body of knowledge on empirical studies on identifying bugs  \cite{10123536} and experimental analysis of security-critical issues in Wasm application development \cite{andre2022developers}. The proposed research has investigated data from social coding and discussion platforms (GitHub, SO) in an attempt to identify, classify, and conceptualize the issues faced by developers and their causes in Wasm application development cycle.
\section{Threats to Validity}
\label{sec:Threats}

\textit{\textbf{External validity}} refers to how generalizable the study's findings are to other contexts or settings related to Wasm issues and causes. One of the possible threat could be missing out some Wasm issues or getting different results from various other platforms/data sources such as GitLab and Bitbucket. In order to minimize this potential biases, we gathered data from two widely-used and popular platforms, namely GitHub and Stack Overflow. These two platforms contains the millions of developers user base. Another potential threat may be not considering all data points for our analysis. To ensure a well-rounded representation of the data, we followed a standard random sampling technique with 95\% confidence level and 5\% margin of error \cite{israel1992determining}.

\textit{\textbf{Internal validity}}  relates to how well a study minimizes bias collection. One of the possible risks includes the qualitative analysis and taxonomy synthesis from the discussions of GitHub and Q\&A posts on Stack Overflow. More specially, the annotation phase could inject subjective bias among the annotators. To mitigate this risk, we conducted a pilot study to establish a shared comprehension of the attributes of Wasm issues. This initial phase also aided in the creation of a robust coding schema for the subsequent annotation process. Furthermore, two authors construct the taxonomies, with a third author conducting a comprehensive validation of the results and resolving any discrepancies through ongoing consensus discussions. Additionally, we calculated Cohen Kappa values to assess the agreement among all authors. Another potential threat to internal validity concerns the selection of open-source GitHub projects. %To address this concern, we established precise criteria for project selection, and once the final set of projects was determined, we conducted manual confirmation to ensure they were indeed Wasm-based applications.
\section{Conclusions}
\label{sec:Conclusion}
In this research, we developed the first-of-its-kind taxonomies for Wasm issues and their causes. \textit{Implications:} This study provides researchers and practitioners with valuable insights into the challenges and complexities involved in the development and deployment of Wasm application. The taxonomy and empirical findings contribute as an evidence-based understanding that is essential for advancing the research in Wasm, which has seen rising attention but still lacks comprehensive issue-related research.

\textit{Needs for future research}: We have three main objectives: (i) To propose a taxonomy of solutions, mapping the relationships among issues, causes, and potential solutions. (ii) To validate the proposed taxonomy of issues, causes, and solutions through an industrial survey, seeking insights from the practitioners' perspective. (iii) To investigate the difficulty and priority levels associated with the identified issues in practical settings.

\section*{Acknowledgments} \label{sec:ack}
This research is funded by Business Finland through the LiquidAI (8542/31/2022) and 6G Soft (8541/31/2022) projects, and by the NSFC China under Grant No. 62172311.
\balance

\bibliographystyle{IEEEtran}
\bibliography{References}

\begin{thebibliography}{10}
\providecommand{\url}[1]{#1}
\csname url@samestyle\endcsname
\providecommand{\newblock}{\relax}
\providecommand{\bibinfo}[2]{#2}
\providecommand{\BIBentrySTDinterwordspacing}{\spaceskip=0pt\relax}
\providecommand{\BIBentryALTinterwordstretchfactor}{4}
\providecommand{\BIBentryALTinterwordspacing}{\spaceskip=\fontdimen2\font plus
\BIBentryALTinterwordstretchfactor\fontdimen3\font minus \fontdimen4\font\relax}
\providecommand{\BIBforeignlanguage}[2]{{%
\expandafter\ifx\csname l@#1\endcsname\relax
\typeout{** WARNING: IEEEtran.bst: No hyphenation pattern has been}%
\typeout{** loaded for the language `#1'. Using the pattern for}%
\typeout{** the default language instead.}%
\else
\language=\csname l@#1\endcsname
\fi
#2}}
\providecommand{\BIBdecl}{\relax}
\BIBdecl

\bibitem{lehmann2020everything}
D.~Lehmann, J.~Kinder, and M.~Pradel, ``Everything old is new again: Binary security of {WebAssembly},'' in \emph{Proceedings of the 29th USENIX Security Symposium (USS)}.\hskip 1em plus 0.5em minus 0.4em\relax USENIX, 2020, pp. 217--234.

\bibitem{haas2017bringing}
A.~Haas, A.~Rossberg, D.~L. Schuff, B.~Titzer, M.~Holman, D.~Gohman, L.~Wagner, A.~Zakai, and J.~Bastien, ``Bringing the web up to speed with {WebAssembly},'' in \emph{Proceedings of the 38th ACM SIGPLAN Conf. on Programming Language Design and Implementation (PLDI)}.\hskip 1em plus 0.5em minus 0.4em\relax ACM, 2017, pp. 185--200.

\bibitem{ketonen2022examining}
T.~Ketonen, ``Examining performance benefits of real-world {WebAssembly} applications: a quantitative multiple-case study,'' Bachelor's Thesis, 2022.

\bibitem{goltzsche2019acctee}
D.~Goltzsche, M.~Nieke, T.~Knauth, and R.~Kapitza, ``Acctee: A {WebAssembly}-based two-way sandbox for trusted resource accounting,'' in \emph{Proceedings of the 20th Int. Middleware Conf. (Middleware)}.\hskip 1em plus 0.5em minus 0.4em\relax ACM, 2019, pp. 123--135.

\bibitem{kotilainen2023webassembly}
P.~Kotilainen, V.~J{\"a}rvinen, J.~Tarkkanen, T.~Autto, T.~Das, M.~Waseem, and T.~Mikkonen, ``{WebAssembly} in iot: Beyond toy examples,'' in \emph{Proceedings of the 23rd Int. Conf. on Web Engineering (ICWE)}.\hskip 1em plus 0.5em minus 0.4em\relax Springer, 2023, pp. 93--100.

\bibitem{kotilainen2022proposing}
P.~Kotilainen, T.~Autto, V.~J{\"a}rvinen, T.~Das, and J.~Tarkkanen, ``Proposing isomorphic microservices based architecture for heterogeneous iot environments,'' in \emph{Proceedings of the 23rd Int. Conf. on Product-Focused Software Process Improvement (PROFES)}.\hskip 1em plus 0.5em minus 0.4em\relax Springer, 2022, pp. 621--627.

\bibitem{herrera2018webassembly}
D.~Herrera, H.~Chen, E.~Lavoie, and L.~Hendren, ``{WebAssembly} and javascript challenge: Numerical program performance using modern browser technologies and devices,'' \emph{University of McGill, Montreal: QC, Technical report SABLE-TR-2018-2}, 2018.

\bibitem{waseem2021nature}
M.~Waseem, P.~Liang, M.~Shahin, A.~Ahmad, and A.~R. Nassab, ``On the nature of issues in five open source microservices systems: An empirical study,'' in \emph{Proceedings of the 25th Int. Conf. on Evaluation and Assessment in Software Engineering (EASE)}.\hskip 1em plus 0.5em minus 0.4em\relax ACM, 2021, pp. 201--210.

\bibitem{waseem2023understanding}
M.~Waseem, P.~Liang, A.~Ahmad, A.~A. Khan, M.~Shahin, P.~Abrahamsson, A.~R. Nasab, and T.~Mikkonen, ``Understanding the issues, their causes and solutions in microservices systems: An empirical study,'' \emph{arXiv preprint arXiv:2302.01894}, 2023.

\bibitem{replpack}
M.~Waseem, T.~Das, A.~Ahmad, P.~Liang, and T.~Mikkonen, ``{Dataset for the Paper: Issues and Their Causes in {WebAssembly} Applications: An Empirical Study},'' \url{https://zenodo.org/record/10528608}, Jan. 2024.

\bibitem{israel1992determining}
G.~D. Israel, ``Determining sample size,'' Florida Cooperative Extension Service, Institute of Food and Agricultural Sciences, University of Florida, Florida, U.S.A, Fact Sheet PEOD-6, November 1992.

\bibitem{cruzes2011recommended}
D.~S. Cruzes and T.~Dyba, ``Recommended steps for thematic synthesis in software engineering,'' in \emph{Proceedings of the 5th ACM/IEEE Int. Symposium on Empirical Software Engineering and Measurement (ESEM)}.\hskip 1em plus 0.5em minus 0.4em\relax IEEE, 2011, pp. 275--284.

\bibitem{li2023warnings}
R.~Li, P.~Liang, and P.~Avgeriou, ``Warnings: Violation symptoms indicating architecture erosion,'' \emph{Information and Software Technology}, vol. 164, p. 107319, 2023.

\bibitem{li2018cid}
L.~Li, T.~F. Bissyand{\'e}, H.~Wang, and J.~Klein, ``Cid: Automating the detection of api-related compatibility issues in android apps,'' in \emph{Proceedings of the 27th ACM SIGSOFT Int. Symposium on Software Testing and Analysis (ISSTA)}.\hskip 1em plus 0.5em minus 0.4em\relax ACM, 2018, pp. 153--163.

\bibitem{eleskovic2020closer}
D.~Eleskovic, ``A closer look at {WebAssembly},'' Bachelor's Thesis, 2020.

\bibitem{bosshard2020use}
B.~Bosshard, ``On the use of web assembly in a serverless context,'' in \emph{Proceedings of the 21st Int. Conf. on Agile Software Development (XP) Workshops}.\hskip 1em plus 0.5em minus 0.4em\relax Springer, 2020, pp. 141--145.

\bibitem{ray2023overview}
P.~P. Ray, ``An overview of {WebAssembly} for iot: Background, tools, state-of-the-art, challenges, and future directions,'' \emph{Future Internet}, vol.~15, no.~8, p. 275, 2023.

\bibitem{vsipek2021next}
M.~{\v{S}}ipek, D.~Muharemagi{\'c}, B.~Mihaljevi{\'c}, and A.~Radovan, ``Next-generation web applications with {WebAssembly} and trufflewasm,'' in \emph{Proceedings of the 44th Int. Convention on Information, Communication and Electronic Technology (MIPRO)}.\hskip 1em plus 0.5em minus 0.4em\relax IEEE, 2021, pp. 1695--1700.

\bibitem{wang2021empowering}
W.~Wang, ``Empowering web applications with {WebAssembly}: are we there yet?'' in \emph{Proceedings of the 36th IEEE/ACM Int. Conf. on Automated Software Engineering (ASE)}.\hskip 1em plus 0.5em minus 0.4em\relax IEEE, 2021, pp. 1301--1305.

\bibitem{niessen2020insights}
T.~Nie{\ss}en, M.~Dawson, P.~Patros, and K.~B. Kent, ``Insights into {WebAssembly}: compilation performance and shared code caching in node.js,'' in \emph{Proceedings of the 30th Annual Int. Conf. on Computer Science and Software Engineering (CASCON)}.\hskip 1em plus 0.5em minus 0.4em\relax ACM, 2020, pp. 163--172.

\bibitem{infoworld2023}
\BIBentryALTinterwordspacing
P.~Krill. (2023) Direct {WebAssembly} compilation comes to rust language. [Online]. Available: \url{https://www.infoworld.com}
\BIBentrySTDinterwordspacing

\bibitem{plainenglish2023}
\BIBentryALTinterwordspacing
H.~Patel. (2023) {WebAssembly}: Unlocking performance and portability for web applications. [Online]. Available: \url{https://javascript.plainenglish.io}
\BIBentrySTDinterwordspacing

\bibitem{cloudflare2023}
\BIBentryALTinterwordspacing
C.~Popoviciu. (2023) Use the language of your choice with pages functions via {WebAssembly}. [Online]. Available: \url{https://blog.cloudflare.com}
\BIBentrySTDinterwordspacing

\bibitem{stievenart2021security}
Q.~Sti{\'e}venart, C.~De~Roover, and M.~Ghafari, ``The security risk of lacking compiler protection in {WebAssembly},'' in \emph{Proceedings of the 21st IEEE Int. Conf. on Software Quality, Reliability and Security (QRS)}.\hskip 1em plus 0.5em minus 0.4em\relax IEEE, 2021, pp. 132--139.

\bibitem{cabrera2020superoptimization}
J.~Cabrera~Arteaga, S.~Donde, J.~Gu, O.~Floros, L.~Satabin, B.~Baudry, and M.~Monperrus, ``Superoptimization of {WebAssembly} bytecode,'' in \emph{Proceedings of the 4th Int. Conf. on Art, Science, and Engineering of Programming (PROGRAMMING): Companion}.\hskip 1em plus 0.5em minus 0.4em\relax ACM, 2020, pp. 36--40.

\bibitem{jangda2019not}
A.~Jangda, B.~Powers, E.~D. Berger, and A.~Guha, ``Not so fast: Analyzing the performance of {WebAssembly} vs. native code,'' in \emph{Proceedings of the USENIX Annual Technical Conf. (ATC)}.\hskip 1em plus 0.5em minus 0.4em\relax USENIX, 2019, pp. 107--120.

\bibitem{chen2022wasai}
W.~Chen, Z.~Sun, H.~Wang, X.~Luo, H.~Cai, and L.~Wu, ``Wasai: uncovering vulnerabilities in wasm smart contracts,'' in \emph{Proceedings of the 31st ACM SIGSOFT Int. Symposium on Software Testing and Analysis (ISSTA)}.\hskip 1em plus 0.5em minus 0.4em\relax ACM, 2022, pp. 703--715.

\bibitem{quan2019evulhunter}
L.~Quan, L.~Wu, and H.~Wang, ``Evulhunter: Detecting fake transfer vulnerabilities for eosio's smart contracts at {WebAssembly}-level,'' \emph{arXiv preprint arXiv:1906.10362}, 2019.

\bibitem{zhou2023wasmod}
J.~Zhou and T.~Chen, ``Wasmod: Detecting vulnerabilities in wasm smart contracts,'' \emph{IET Blockchain}, 2023.

\bibitem{9678776}
A.~Romano, X.~Liu, Y.~Kwon, and W.~Wang, ``An empirical study of bugs in {WebAssembly} compilers,'' in \emph{Proceedings of the 36th IEEE/ACM Int. Conf. on Automated Software Engineering (ASE)}.\hskip 1em plus 0.5em minus 0.4em\relax IEEE, 2021, pp. 42--54.

\bibitem{10123536}
Y.~Wang, Z.~Zhou, Z.~Ren, D.~Liu, and H.~Jiang, ``A comprehensive study of {WebAssembly} runtime bugs,'' in \emph{Proceedings of the 30th IEEE Int. Conf. on Software Analysis, Evolution and Reengineering (SANER)}.\hskip 1em plus 0.5em minus 0.4em\relax IEEE, 2023, pp. 355--366.

\bibitem{yixuan2023characterizing}
Y.~Zhang, S.~Cao, H.~Wang, Z.~Chen, X.~Luo, D.~Mu, Y.~Ma, G.~Huang, and X.~Liu, ``Characterizing and detecting {WebAssembly} runtime bugs,'' \emph{ACM Transactions on Software Engineering and Methodology}, 2023.

\bibitem{yan2021understanding}
Y.~Yan, T.~Tu, L.~Zhao, Y.~Zhou, and W.~Wang, ``Understanding the performance of {WebAssembly} applications,'' in \emph{Proceedings of the 21st ACM Internet Measurement Conf. (IMC)}.\hskip 1em plus 0.5em minus 0.4em\relax ACM, 2021, pp. 533--549.

\bibitem{andre2022developers}
P.~M. Andr{\'e}, Q.~Sti{\'e}venart, and M.~Ghafari, ``Developers struggle with authentication in blazor {WebAssembly},'' in \emph{Proceedings of the 38th IEEE Int. Conf. on Software Maintenance and Evolution (ICSME)}.\hskip 1em plus 0.5em minus 0.4em\relax IEEE, 2022, pp. 389--393.

\end{thebibliography}

\end{document}